\documentclass[10pt]{article}
\usepackage{latexsym,graphicx}
\newcommand{\be}{\begin{equation}}
\newcommand{\ee}{\end{equation}}
\def\n{\noindent}
\catcode `\@=11
\catcode `\@=12
\begin{document}
\begin{center}
{\bf {Bianchi Type-II String Cosmological Models in Normal Gauge for Lyra's Manifold
with Constant Deceleration Parameter}} \\
\vspace{5mm}
\normalsize{Shilpi Agarwal $^{1}$, R. K. Pandey  $^{2}$ and Anirudh Pradhan {$^{3}$}} \\
\vspace{10mm}
\normalsize{$^{1}$ Department of Mathematics, Uttaranchal Institute of Technology, Arcadia Grant, 
Chandanwari, Dehradun 248 007, India \\
e-mail: shilpisinghal77@gmail.com }\\
\vspace{5mm}
\normalsize{$^{2}$ Department of Mathematics, D. B. S. Post-graduate College, Dehradun 248 001, India}\\
\vspace{5mm}
\normalsize{$^{3}$ Department of Mathematics, Hindu Post-graduate College,
Zamania 232 331, Ghazipur, India \\
E-mail: pradhan@iucaa.ernet.in}\\
\end{center}
\vspace{10mm}
\begin{abstract} 
The present study deals with a spatially homogeneous and anisotropic Bianchi-II cosmological models representing 
massive strings in normal gauge for Lyra's manifold by applying the variation law for generalized Hubble's parameter 
that yields a constant value of deceleration parameter. The variation law for Hubble's parameter generates
two types of solutions for the average scale factor, one is of power-law type and other is of the exponential
form. Using these two forms, Einstein's modified field equations are solved separately that correspond to expanding
singular and non-singular models of the universe respectively. The energy-momentum tensor for such string
as formulated by Letelier (1983) is used to construct massive string cosmological models for which we assume
that the expansion ($\theta$) in the model is proportional to the component $\sigma^{1}_{~1}$ of the shear
tensor $\sigma^{j}_{~i}$. This condition leads to $A = (BC)^{m}$, where A, B and C are the metric coefficients
and m is proportionality constant. Our models are in accelerating phase which is consistent to the recent
observations. It has been found that the displacement vector $\beta$ behaves like cosmological term $\Lambda$ in 
the normal gauge treatment and the solutions are consistent with recent observations of SNe Ia. It has been found 
that massive strings dominate in the decelerating universe whereas strings dominate in the accelerating universe. 
Some physical and geometric behaviour of these models are also discussed.
\end{abstract}
\smallskip
\n PACS: {98.80.Cq, 04.20.-q} \\ 
\smallskip
\n Keywords: {String, Bianchi type-II models, Lyra's manifold, Accelerating models}
\section{Introduction}
In Einstein's general theory, the curvature of a space-time is influenced by matter, and provides 
the geometrical description of matter. Einstein (1917) succeeded in geometrizing gravitation by expressing 
gravitational potential in terms of metric tensor. Weyl, in 1918, was inspired by it and he was the the 
first to unify gravitation and electromagnetism in a single space-time geometry. He showed how can one 
introduce a vector field in the Riemannian space-time with an intrinsic geometrical significance. But this 
theory was not accepted as it was based on non-integrability of length transfer. Lyra \cite{ref1} introduced 
a gauge function, i.e., a displacement vector in Riemannian space-time which removes the non-integrability 
condition of a vector under parallel transport. In this way Riemannian geometry was given a new modification 
by him and the modified geometry was named as Lyra's manifold. In consecutive investigations Sen \cite{ref2}, 
Sen and Dunn \cite{ref3}  proposed a new scalar-tensor theory of gravitation and constructed an analog of the 
Einstein field equations based on Lyra's geometry. It is, thus, possible \cite{ref2} to construct a geometrized 
theory of gravitation and electromagnetism much along the lines of Weyl's ``unified'' field theory, however, 
without the inconvenience of non-integrability length transfer. \\

Halford \cite{ref4} has pointed out that the constant vector displacement field $\phi_i$ in Lyra's manifold 
plays the role of cosmological constant $\Lambda$ in the normal general relativistic treatment. It
is shown by Halford \cite{ref5} that the scalar-tensor treatment based on Lyra's geometry predicts the same 
effects, within observational limits as the Einstein's theory. The Sen \cite{ref2} theory and its more 
generalizations (Sen and Dun \cite{ref3}; Sen and Vanstone \cite{ref6}) have received considerable attention 
in cosmological context. Several investigators {\cite{ref6}$-$\cite{ref27}} have studied cosmological models 
based on Lyra's manifold in different contexts. Soleng \cite{ref7} has pointed out that the cosmologies based 
on Lyra's manifold with constant gauge vector $\phi$ will either include a creation  field and be equal to 
Hoyle's creation field cosmology {\cite{ref28}$-$\cite{ref30}} or contain a special vacuum field which together 
with the gauge vector term may be considered as a cosmological term. In the latter case the solutions are equal 
to the general relativistic cosmologies with a cosmological term. \\

In recent years, there has been considerable interest in string cosmology. Cosmic strings are
topologically stable objects which might be found during a phase transition in the early
universe (Kibble \cite{ref31}). Cosmic strings play an important role in the study of the early universe.
These arise during the phase transition after the big bang explosion as the temperature goes down
below some critical temperature as predicted by grand unified theories (Zel'dovich et al. \cite{ref32};
Kibble \cite{ref31,ref33}; Everett \cite{ref34}; Vilenkin \cite{ref35}). It is believed that cosmic strings
give rise to density perturbations which lead to the formation of galaxies (Zel'dovich \cite {ref36}).
These cosmic strings have stress-energy and couple to the gravitational field. Therefore it is interesting
to study the gravitational effects that arise from strings. The pioneering work in the formulation of the
energy-momentum tensor for classical massive strings was done by Letelier \cite{ref37} who considered
the massive strings to be formed by geometric strings with particle attached along its extension.
Letelier \cite{ref38} first used this idea in obtaining cosmological solutions in Bianchi I and
Kantowski-Sachs space-times. Stachel \cite{ref39} has studied massive string. During the last ten years,
many authors (\cite{ref40}$-$\cite{ref60} and references therein) have discussed the string cosmological
models in different contexts.\\

Recently, Pradhan et al. {\cite{ref61}$-$\cite{ref67}}, Casama et al. \cite{ref68}, Bali and  Chandnani 
\cite{ref69,ref70}, Kumar and Singh \cite{ref71}, Ram, Zeyauddin and Singh \cite{ref72}, Singh \cite{ref73} 
and Rao, Vinutha and Santhi \cite{ref74} have studied cosmological models based on Lyra's geometry in various 
contexts. With these motivations, in this paper, we have obtained homogeneous and anisotropic Bianchi type II 
string cosmological models of perfect fluid distribution of matter for the field equations in normal gauge for 
Lyra's manifold where gauge function $\beta$ is taken as time dependent. This paper is organized as follows. 
In Section $1$  the motivation for the present work is discussed. The metric and the field equations are presented 
in Section $2$. In Section $3$, we deal with exact solutions of the field equations with two types of string 
cosmological models using the power-law and exponential-law of expansion of the universe respectively and their 
physical and geometric properties of both models have been described. Finally, in Section $4$ the
concluding remarks have been given.\\
\section{The metric and basic equations}
We consider totally anisotropic Bianchi type-II line element, given by
\begin{equation}
\label{eq1} ds^{2} = - dt^{2} + A^{2}(dx - zdy)^{2} + B^{2} dy^{2} +
C^{2} dz^{2},
\end{equation}
where the metric potentials $A$, $B$ and $C$ are functions of $t$ alone. This ensures that the model
is spatially homogeneous. \\
The energy-momentum tensor for a cloud of massive string with perfect fluid is taken as
\begin{equation}
\label{eq2}
T^{j}_{i} = (\rho + p)v_{i}v^{j} + p g^{j}_{i} - \lambda x_{i}x^{j},
\end{equation}
where $p$ is the isotropic pressure; $\rho$ is the rest energy density for a cloud of strings with particles
attached to them; $\lambda$ is the string tension density; $v^{i}=(0,0,0,1)$ is the four-velocity of the
particles, and $x^{i}$ is a unit space-like vector representing the direction of strings so that $x^{2} =
0 = x^{3} = x^{4}$ and $x^{1} \ne 0$. The vectors $v^{i}$ and $x^{i}$ satisfy the conditions
\begin{equation}
\label{eq3}
v_{i}v^{i} = -x_{i}x^{i} = -1,\;\; v^{i}x_{i}=0.
\end{equation}
Choosing $x^{i}$ parallel to $\partial/\partial x$, we have
\begin{equation}
\label{eq4}
x^{i} = (A^{-1},0,0,0).
\end{equation}
If the particle density of the configuration is denoted by $\rho_{p}$, then
\begin{equation}
\label{eq5}
\rho = \rho_{p} + \lambda.
\end{equation}
The field equations (in gravitational units $c = 1$, $8\pi G = 1$), in normal gauge for Lyra's manifold, obtained by
Sen \cite{ref4} as
\begin{equation}
\label{eq6}
R^{j}_{i} - \frac{1}{2} g^{j}_{i} R + \frac{3}{2} \phi_i \phi^{j} - \frac{3}{4} g^{j}_{i} \phi_k \phi^k = 
- T^{j}_{i},
\end{equation}
where $\phi_{i}$ is the displacement field vector defined as
\begin{equation}
\label{eq7} \phi_{i} = (0, 0, 0, \beta(t)),
\end{equation}
and other symbols have their usual meaning as in Riemannian geometry. \\\\
In a co-moving co-ordinate system, the Einstein's modified field equation (\ref{eq6}) with (\ref{eq2}) for the 
metric (\ref{eq1}) subsequently lead to the following system of equations:
\begin{equation}
\label{eq8} \frac{\ddot{B}}{B} + \frac{\ddot{C}}{C} + \frac{\dot{B}\dot{C}}{BC} - \frac{3}{4}\frac{A^{2}}{B^{2}C^{2}} 
+ \frac{3}{4}\beta^{2}  = - p + \lambda,
\end{equation}
\begin{equation}
\label{eq9} \frac{\ddot{C}}{C} + \frac{\ddot{A}}{A} + \frac{\dot{C}\dot{A}}{CA} + \frac{1}{4}\frac{A^{2}}
{B^{2}C^{2}} + \frac{3}{4}\beta^{2} = - p,
\end{equation}
\begin{equation}
\label{eq10} \frac{\ddot{A}}{A} + \frac{\ddot{B}}{B} + \frac{\dot{A}\dot{B}}{AB} + \frac{1}{4}\frac{A^{2}}
{B^{2}C^{2}} + \frac{3}{4}\beta^{2} = - p,
\end{equation}
\begin{equation}
\label{eq11} \frac{\dot{A}\dot{B}}{AB} + \frac{\dot{B}\dot{C}}{BC} + \frac{\dot{C}\dot{A}}{CA} -
\frac{1}{4}\frac{A^{2}}{B^{2}C^{2}} + \frac{3}{4}\beta^{2} =  \rho.
\end{equation}
Here, and in what follows, a dot indicates ordinary differentiation with respect to $t$. 
The energy conservation equation $T^{i}_{i;j} = 0$ leads to
\begin{equation}
\label{eq12}  \dot\rho + (\rho + p)\left(\frac{\dot{A}}{A} + \frac{\dot{B}}{B} + \frac{\dot{C}}{C}\right)
- \lambda\frac{\dot{A}}{A} = 0\;,
\end{equation}
and conservation of R. H. S. of Eq. (\ref{eq6}) leads to
\begin{equation}
\label{eq13} \left(R^{j}_{i} - \frac{1}{2} g^{j}_{i} R\right)_{;j} + \frac{3}{2}\left(\phi_i \phi^{j}\right)_{;j} - 
\frac{3}{4} \left(g^{j}_{i} \phi_k \phi^k \right)_{;j} = 0.
\end{equation}
Equation (\ref{eq13}) reduces to
\[
\frac{3}{2}\phi_{i}\left[\frac{\partial{{\phi^{j}}}}{\partial{x^{j}}} + \phi^{l}\Gamma^{j}_{lj}\right] +
\frac{3}{2}\phi^{j}\left[\frac{\partial{{\phi_{i}}}}{\partial{x^{j}}} - \phi_{l}\Gamma^{l}_{ij}\right] -
\frac{3}{4}g^{j}_{i}\phi_{k}\left[\frac{\partial{{\phi^{k}}}}{\partial{x^{j}}} + \phi^{l}\Gamma^{k}_{lj}\right] -
\]
\begin{equation}
\label{eq14}
\frac{3}{4}g^{j}_{i}\phi^{k}\left[\frac{\partial{{\phi_{k}}}}{\partial{x^{j}}} - \phi_{l}\Gamma^{l}_{kj}\right] = 0.
\end{equation}
Equation (\ref{eq14}) is identically satisfied for $i = 1, 2, 3$.
For $i = 4$, Eq. (\ref{eq14}) reduces to
\[
\frac{3}{2}\beta\left[\frac{\partial{(g^{44}\phi_{4})}}{\partial{x^{4}}} + \phi^{4}\Gamma^{4}_{44}\right] +
\frac{3}{2}g^{44}\phi_{4}\left[\frac{\partial{{\phi_{4}}}}{\partial{t}} - \phi_{4}\Gamma^{4}_{44}\right] -
\frac{3}{4}g^{4}_{4}\phi_{4}\left[\frac{\partial{{\phi^{4}}}}{\partial{x^{4}}} + \phi^{4}\Gamma^{4}_{44}\right] -
\]
\begin{equation}
\label{eq15}
\frac{3}{4}g^{4}_{4}g^{44}\phi^{4}\left[\frac{\partial{{\phi_{4}}}}{\partial{t}} - \phi^{4}\Gamma^{4}_{44}\right] = 0.
\end{equation}
which leads to
\begin{equation}
\label{eq16}\frac{3}{2}\beta\dot{\beta} + \frac{3}{2}\beta^{2}\left(\frac{\dot{A}}{A} + \frac{\dot{B}}{B} + 
\frac{\dot{C}}{C}\right) = 0.
\end{equation}
Thus, equation (\ref{eq12}) combined with (\ref{eq16}) is the resulting equation when energy conservation 
equation is satisfied in the given system. It is important to mention here that the conservation equation 
in Lyra's manifold is not satisfied as in general relativity. Actually, conservation equation in Lyra's manifold 
is satisfied only on giving some special condition on displacement vector $\beta$ as shown above.
\section{Solutions of the field equations}
Equations (\ref{eq8})-(\ref{eq11}) and (\ref{eq16}) are five equations in seven unknown parameters $A$, $B$, $C$, 
$p$, $\rho$, $\lambda$ and $\beta$. Two additional constraints relating these parameters are required to obtain 
explicit solutions of the system. We first assume that the component $\sigma^{1}_{~1}$ of the shear tensor
$\sigma^{j}_{~i}$ is proportional to the expansion scalar ($\theta$). This condition leads to the
following relation between the metric potentials:
\begin{equation}
\label{eq17} A = (BC)^{m},
\end{equation}
where $m$ is a positive constant. The motive behind assuming this condition is explained with
reference to Thorne \cite{ref75}, the observations of the velocity-red-shift relation for extragalactic
sources suggest that Hubble expansion of the universe is isotropic today within $\approx 30$ per cent
\cite{ref76,ref77}. To put more precisely, red-shift studies place the limit
$$
\frac{\sigma}{H} \leq 0.3
$$
on the ratio of shear $\sigma$ to Hubble constant $H$ in the neighbourhood of our Galaxy today. Collins
et al. \cite{ref78} have pointed out that for spatially homogeneous metric, the normal congruence to the
homogeneous expansion satisfies that the condition $\frac{\sigma}{\theta}$ is constant. \\\\\\
Considering $(ABC)^{\frac{1}{3}}$ as the average scale factor of the anisotropic Bianchi-II space-time,
the Hubble parameter may be written as
\begin{equation}
\label{eq18} H = \frac{1}{3}\left(\frac{\dot{A}}{A} + \frac{\dot{B}}{B} + \frac{\dot{C}}{C}\right).
\end{equation}
Secondly, we utilize the special law of variation for the Hubble parameter given by Berman \cite{ref79},
which yields a constant value of deceleration parameter. Here, the law reads as
\begin{equation}
\label{eq19} H = \ell (ABC)^{-\frac{n}{3}},
\end{equation}
where  $\ell > 0$  and  $n \geq 0$ are constants. Such type of relations have already been considered
by Berman and Gomide \cite{ref80} for solving FRW models. Latter on many authors
(see, Saha and Rikhvitsky \cite{ref81}, Saha \cite{ref82}, Singh et al. \cite{ref83},
Singh and Chaubey \cite{ref84,ref85}, Zeyauddin and Ram \cite{ref86}, Singh and Baghel \cite{ref87},
Pradhan and Jotania \cite{ref88} and references therein) have studied flat FRW and Bianchi type models
by using the special law for Hubble parameter that yields constant value of deceleration parameter. \\\\
From equations (\ref{eq18}) and (\ref{eq19}), we get
\begin{equation}
\label{eq20} \frac{1}{3}\left(\frac{\dot{A}}{A} + \frac{\dot{B}}{B} + \frac{\dot{C}}{C}\right) =
\ell (ABC)^{-\frac{n}{3}}.
\end{equation}
Integration of (\ref{eq20}) gives
\begin{equation}
\label{eq21} ABC = (n \ell t + c_{1})^{\frac{3}{n}}, \;\;\;\;(n\neq 0)
\end{equation}
\begin{equation}
\label{eq22} ABC = c_{2}^{3}e^{3 \ell t},\;\;\;\;(n=0)
\end{equation}
where $c_{1}$ and $c_{2}$ are constants of integration. Thus, the law (\ref{eq19}) provides power-law
(\ref{eq21}) and exponential-law (\ref{eq22}) of expansion of the universe. \\\\
The value of deceleration parameter ($q$), is then found to be
\begin{equation}
\label{eq23} q = n - 1,
\end{equation}
which is a constant. The sign of $q$ indicated whether the model inflates or not. The positive sign of
$q$  (i.e. $n > 1$) correspond to ``standard" decelerating model whereas the negative sign of $q$
(i.e. $0 \leq n < 1$) indicates inflation. It is remarkable to mention here that though the current observations of
SNe Ia and CMBR favours accelerating models, but both do not altogether rule out the
decelerating ones which are also consistent with these observations (see, Vishwakarma \cite{ref89}). \\

Subtracting (\ref{eq9}) from (\ref{eq10}), and taking integral of the resulting equation two times, we get
\begin{equation}
\label{eq24}
\frac{B}{C} = c_{2} \exp \left[c_{3} \int (ABC)^{-1} dt \right],
\end{equation}
where  $\;c_{2} $ and  $\;c_{3} $ are constants of integration. \\

In the following subsections, we discuss the string cosmology in normal gauge for Lyra's manifold by using the 
power-law (\ref{eq21}) and exponential-law (\ref{eq22}) of expansion of the universe.
\begin{figure}[ht]
\centering
\includegraphics[width=8cm,height=8cm,angle=0]{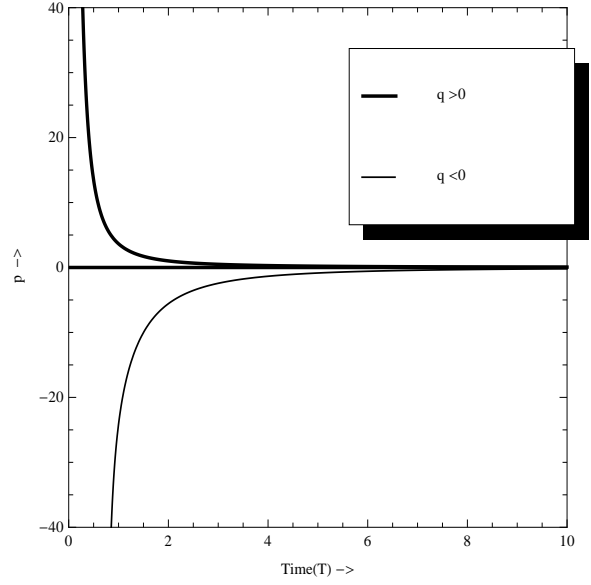} \\
\caption{The plots of isotropic pressure $p$ versus $T$ in power-law expansion for $q > 0$ and $q < 0$.} 
\end{figure}
\begin{figure}[ht]
\centering
\includegraphics[width=8cm,height=8cm,angle=0]{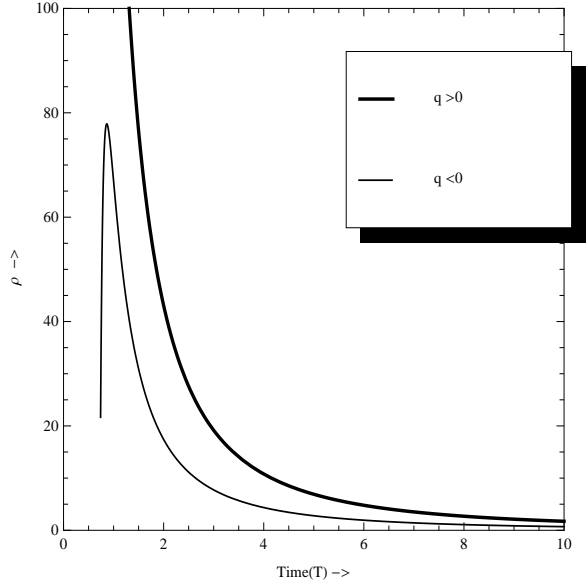} \\
\caption{The plots of proper energy density $\rho$ versus $T$ in power-law expansion for $q > 0$ and $q < 0$.}
\end{figure} 
\begin{figure}[ht]
\centering
\includegraphics[width=8cm,height=8cm,angle=0]{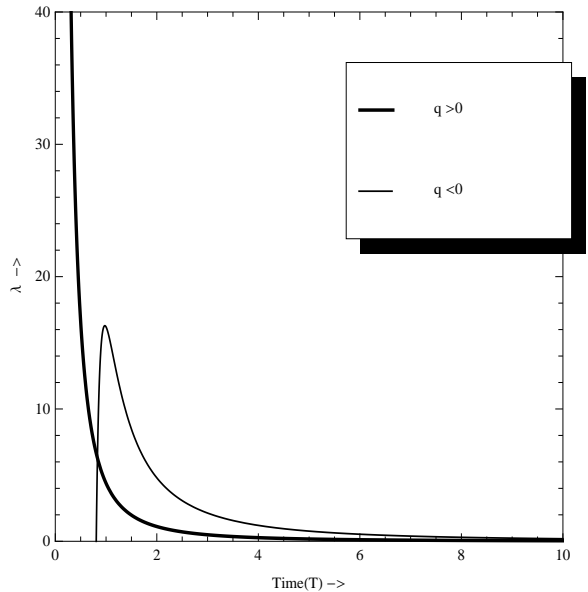} \\
\caption{The plots of string tension density $\lambda$ versus $T$ in powe-law expansion for $q > 0$ and $q < 0$.}
\end{figure}
\begin{figure}[ht]
\centering
\includegraphics[width=8cm,height=8cm,angle=0]{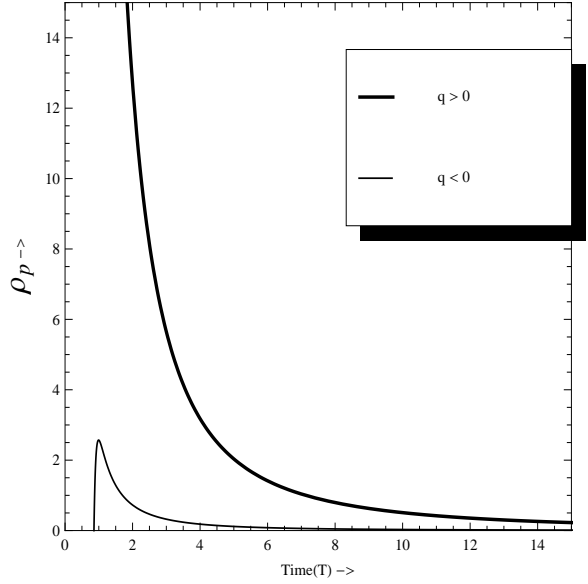} \\
\caption{The plots of particle density $\rho_{p}$ versus $T$ in power-law expansion for $q > 0$ and $q < 0$.}
\end{figure}
\begin{figure}[ht]
\centering
\includegraphics[width=8cm,height=8cm,angle=0]{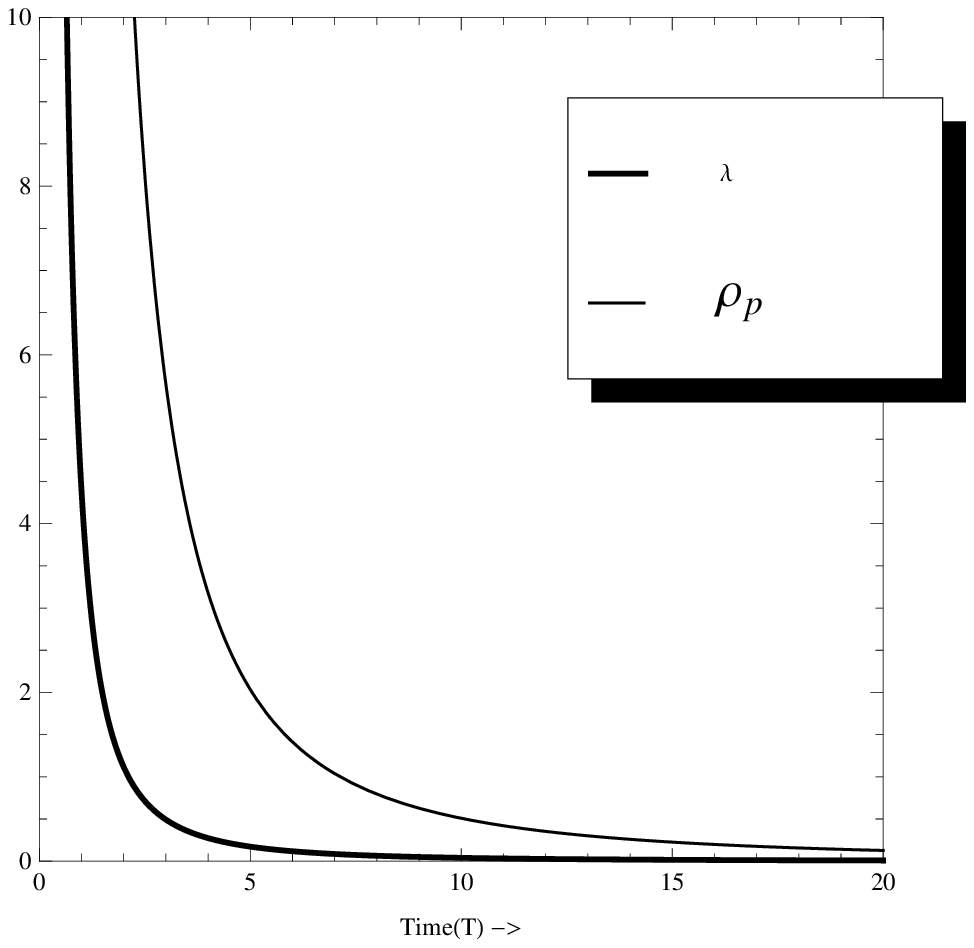} \\
\caption{The plots of $\rho_{p}$ and $\lambda$ versus $T$ in power-law expansion for $q > 0$.}
\end{figure}
\begin{figure}[ht]
\centering
\includegraphics[width=8cm,height=8cm,angle=0]{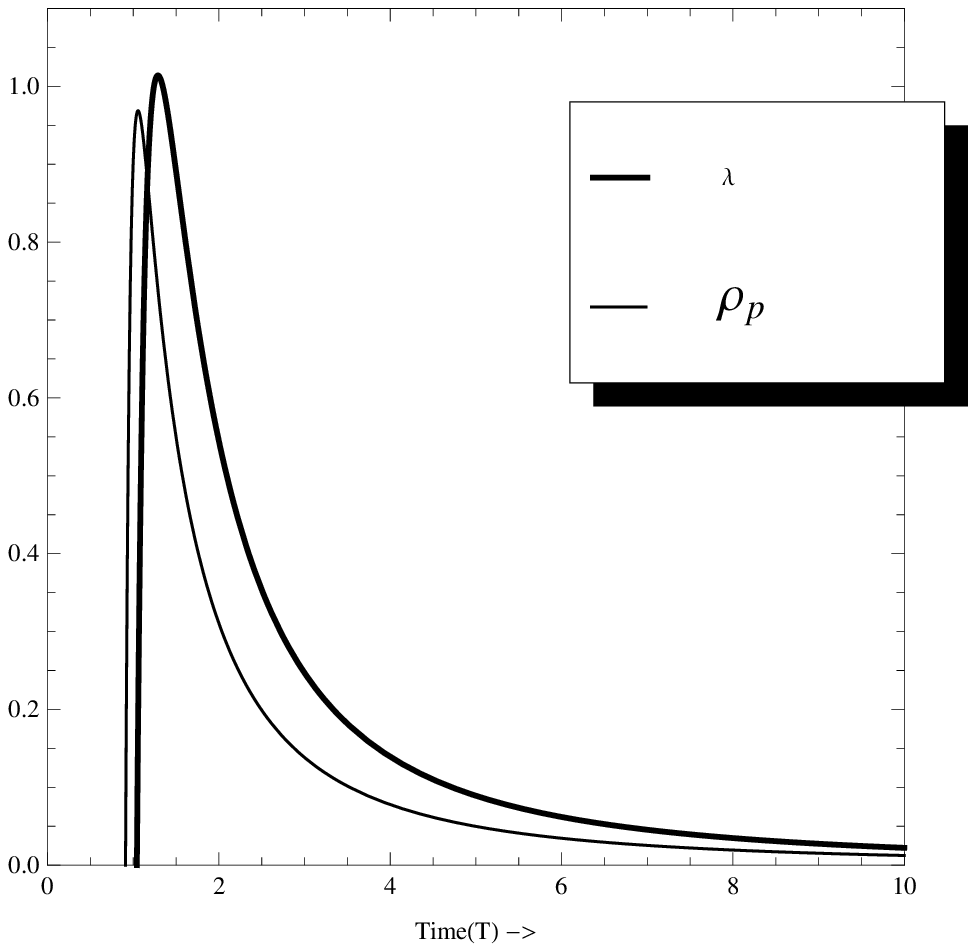} \\
\caption{The plots of $\rho_{p}$ and $\lambda$ versus $T$ in power-law expansion for $q < 0$.}
\end{figure}

\subsection{String Cosmology with Power-law}
Solving the equations (\ref{eq17}), (\ref{eq21}) and (\ref{eq24}), we obtain the metric functions as
\begin{equation}
\label{eq25} A(t) = (n \ell t + c_{1})^{\frac{3m}{n(m + 1)}} \;,
\end{equation}
\begin{equation}
\label{eq26} B(t) = \sqrt{c_{2} }(n \ell t + c_{1})^{\frac{3}{2n(m + 1)}} \exp{\left[\frac{c_{3} }{2 \ell (n-3)}
(n\ell t + c_{1})^{\frac{n - 3}{n} }\right]} \;,
\end{equation}
\begin{equation}
\label{eq27} C(t) = \frac{1}{\sqrt{c_{2}}}(n \ell t + c_{1})^{\frac{3}{2n(m + 1)}}\exp{\left[-\frac{c_{3} }
{2 \ell (n - 3)} (n \ell t + c_{1})^{\frac{n - 3}{n} }\right]} \;,
\end{equation}
provided $\;n\neq3$. \\\\
Therefore the metric (\ref{eq1}) reduces to
\[
 ds^{2} = -dt^{2} + (n\ell t + c_{1})^{\frac{6m}{n(m + 1)}}(dx - zdy)^{2} +
\]
\[
 c_{2}(n \ell t + c_{1})^{\frac{3}{n(m + 1)}} \exp{\left[\frac{c_{3} }{\ell (n-3)}
(n \ell t + c_{1})^{\frac{n - 3}{n} }\right]}dy^{2} +
\]
\begin{equation}
\label{eq28}
\frac{1}{c_{2}}(n \ell t + c_{1})^{\frac{3}{n(m + 1)}}\exp{\left[-\frac{c_{3} }
{\ell (n - 3)} (n \ell t + c_{1})^{\frac{n - 3}{n} }\right]}dz^{2}.
\end{equation}
By using the transformation $x = X$, $y = Y$, $z = Z$, $n\ell t + c_{1} = T$, the space-time (\ref{eq28})
is reduced to
\[
 ds^{2} = -\frac{dT^{2}}{(n \ell)^{2}} + T^{\frac{6m}{n(m + 1)}}dX^{2} +
\]
\[
 c_{2}T^{\frac{3}{n(m + 1)}} \exp{[\frac{c_{3}}{\ell(n - 3)}T^{\frac{n - 3}{n}}]}dY^{2}
\]
\begin{equation}
\label{eq29}
+ \frac{1}{c_{2}}T^{\frac{3}{n(m + 1)}} \exp{[- \frac{c_{3}}{\ell(n - 3)}T^{\frac{n - 3}{n}}]} dZ^{2}.
\end{equation}
Eq. (\ref{eq16}) gives either $\beta = 0$ or $\frac{3}{2} \dot{\beta} + \frac{3}{2} \beta \left(\frac{\dot{A}}{A} 
+ \frac{\dot{B}}{B} + \frac{\dot{C}}{C}\right) = 0$.
Therefore
\begin{equation}
\label{eq30} \frac{\dot{\beta}}{\beta}   = - \left(\frac{\dot{A}}{A} + \frac{\dot{B}}{B} + 
\frac{\dot{C}}{C}\right),
\end{equation}
which reduces to
\begin{equation}
\label{eq31} \frac{\dot{\beta}}{\beta} = - \frac{3}{nT}.
\end{equation}
Integrating Eq. (\ref{eq31}), we obtain
\begin{equation}
\label{eq32} \beta = kT^{-\frac{3}{n}},
\end{equation}
where $k$ is an integrating constant. \\\\
Halford \cite{ref4} has pointed out that the displacement field $\phi_i$ in Lyra's manifold plays the role of 
cosmological constant $\Lambda$ in the normal general relativistic treatment. From Eq. (\ref{eq32}), it is observed 
that for $k > 0$ and $n > 0$, the displacement vector $\beta(T)$ is a decreasing function of time and it approaches 
to a small positive value at late time (i.e. present epoch), which is corroborated with Halford as well as with 
the recent observations of SNe Ia. Recent cosmological observations of SNe Ia (Garnavich et al. \cite{ref90,ref91}; 
Perlmutter et al. \cite{ref92}$-$\cite{ref94}; Riess et al. \cite{ref95,ref96}; Schmidt et al. \cite{ref97}) suggest 
the existence of a positive cosmological constant $\Lambda$ with the magnitude $\Lambda(G\hbar/c^{3})\approx 10^{-123}$. 
These observations on magnitude and red-shift of type Ia supernova suggest that our universe may be an accelerating
one with induced cosmological density through the cosmological $\Lambda$-term. But this does not rule out the
decelerating ones which are also consistent with these observations (Vishwakarma \cite{ref89}). Thus the nature of
$\beta(T)$ in our derived model is supported by recent observations. \\\\
The expressions for the isotropic pressure ($p$), the proper energy density ($\rho$), the string tension
($\lambda$) and the particle density ($\rho_{p}$) for the model (\ref{eq29}) are obtained as
\[
p = \frac{3\ell^{2}[4m^{2}(n - 3) + 6m(n - 1) + 2n - 3]}{4(m + 1)^{2}}T^{-2} -
\]
\begin{equation}
\label{eq33}
\frac{c_{3}^{2}}{4}T^{-\frac{6}{n}} - \frac{1}{4}T^{\frac{6(m - 1)}{n (m + 1)}}
- \frac{3}{4}k^{2}T^{-\frac{6}{n}},
\end{equation}
\begin{equation}
\label{eq34}
\rho = \frac{9\ell^{2}(4m + 1)}{4(m + 1)^{2}}T^{-2} - \frac{c_{3}^{2}}{4}T^
{-\frac{6}{n}} - \frac{1}{4}T^{\frac{6(m - 1)}{n (m + 1)}} + \frac{3}{4}k^{2}T^{-\frac{6}{n}},
\end{equation}
\begin{equation}
\label{eq35}
\lambda = \frac{3\ell^{2}(2m - 1)(n - 3)}{2(m + 1)}T^{-2} - T^{\frac{6(m - 1)}{n (m + 1)}},
\end{equation}
\[
\rho_{p} = \frac{3\ell^{2}[3(4m + 1) - 2(m + 1)(2m - 1)(n - 3)]}{4(m + 1)^{2}}T^{-2}
\]
\begin{equation}
\label{eq36}
- \frac{c_{3}^{2}}{4}T^{-\frac{6}{n}} + \frac{3}{4}T^{\frac{6(m - 1)}
{n (m + 1)}} + \frac{3}{4}k^{2}T^{-\frac{6}{n}}.
\end{equation}
It is evident that the energy conditions  $\rho \geq 0$ and $\rho_{p} \geq 0$ are satisfied under the appropriate 
choice of constants. We observe that all the parameters diverge at $T = 0$. Therefore, the model has a singularity
at $T = 0$. This singularity is of Point Type (MacCallum \cite{ref98}) since all the scale factors diverge at 
$T = 0$. The cosmological evolution of B-II space-time is expansionary, with all the three scale factors 
monotonically increasing function of time. So, the universe starts expanding with a big bang singularity in the 
derived model. The parameters $p$, $\rho$, $\rho_{p}$ and $\lambda$ start off with extremely large values, 
which continue to decrease with the expansion of the universe provided $m < 1$. In particular, the large values of 
$\rho_{p}$ and $\lambda$ in the beginning suggest that strings dominate the early universe. For sufficiently large 
times, the $\rho_{p}$ and $\lambda$ become negligible. Therefore, the strings disappear from the universe for large
times (at later stage of evolution). That is why, the strings are not observable in the present universe. \\\\
For $ n < 1$, the model is accelerating whereas for $n > 1$ it goes to decelerating phase. In what follows, we
compare the two modes of evolution through graphical analysis of various parameters. We have chosen $n = 2$,
{\it i.e.} $q = 1$ to describe the decelerating phase while the accelerating mode has been accounted by choosing
$n = 0.4$, {\it i.e.}, $q = -0.6$. The other constants are chosen as $\ell = 2$, $K = 1$, $c_{3} = 1$, $m = 0.3$.
\\\\
{\bf Figure 1} shows the plots of the variation of pressure versus time in the two modes of evolution of the 
universe (i.e for decelerating and accelerating respectively). We observe that the pressure is positive in the 
decelerating phase which decreases with the evolution of the universe. But in the accelerating phase, the pressure 
is negative rises in the early phase and then it increases with time as aspected. The pressure in both phases are 
always  becomes negligible at late time. \\\\
From Eq. (\ref{eq34}), it is observed that the rest energy density $\rho$ is a decreasing function of time and
$\rho > 0$ always. The rest energy density has been plotted versus time in {\bf Figure 2} for $q > 0$ and $q < 0$ 
respectively. It is evident that the rest energy density remains positive in both modes of evolution. However, 
it decreases more sharply with the cosmic time in the decelerating universe compare to accelerating universe. \\\\
From Eq. (\ref{eq35}), it is observed that the tension density $\lambda$ is a decreasing function of time and
$\lambda > 0$ always. We have graphed the string tension density versus time in the {\bf Figure 3} in both decelerating
and accelerating modes of the universe. It is evident that the $\lambda$ remains positive in both modes of
evolution. However, it decreases more sharply with the cosmic time in the accelerating universe compare to
decelerating universe. In the early phase of universe, the string tension density of both modes will dominate
the dynamics and later time it approaches to zero. It is worth mentioning that string tension density is less
in decelerating phase compare to accelerating phase and due to this in decelerating phase the massive string
disappear from the evolution phase of the universe at later stage (i.e. present epoch). \\\\
From Eq. (\ref{eq36}), it is evident that the particle density $\rho_{p}$ is a decreasing function of time and
$\rho_{p} > 0$ for all time. {\bf Figure 4} shows the plots of particle density versus time in both decelerating
and accelerating modes of the universe. Here it is to be noted that $\rho_{p}$ in the decelerating phase is
more larger than the accelerating phase throughout the evolution of the universe. In the accelerating phase $\rho_{p}$
increases rapidly in initial stage, it attains maximum value at some epoch closer to the early phase of the
universe. In the later stage, it decreases from its maximum value with time and approaches to small value at
late time. \\\\
{\bf Figure 5} shows the comparative behaviour of particle energy density and string tension density versus time in
the decelerating mode. It is observed that $\rho_{p} > \lambda$, i.e., particle energy density remains larger
than the string tension density during the cosmic expansion (see, Refs. Kibble \cite{ref31}; Krori et al.
\cite{ref99}), especially in early universe. This shows that massive strings dominate the early universe
evolving with deceleration  and in later phase  will disappear which is in agreement with current astronomical
observations.\\\\
{\bf Figure 6} demonstrates the variation of $\rho_{p}$ and $\lambda$ versus the cosmic time for accelerating
mode of the evolution. In this case, we observe that  $\rho_{p} < \lambda$. Therefore, according to Kibble 
\cite{ref31} and Krori et al. \cite{ref99}, strings dominate the universe evolving with acceleration. If this 
is so, we should have some signature of massive string at present epoch of the observations. However, it is not
been seen so far. \\\\
It follows that the dynamics of the strings depends on the value of n or q. Further, it is observed that for
sufficiently large times, the $\rho_{p}$ and $\lambda$ tend to zero. Therefore, the strings disappear from
the universe at late time (i.e. present epoch). The same is predicted by the current observations.\\\\
The rate of expansion $H_{i}$ in the direction of x, y and z are given by
\begin{equation}
\label{eq37} H_{x} = \frac{\dot{A}}{A} = \frac{3m\ell}{m+1} {T}^{-1}\; ,
\end{equation}
\begin{equation}
\label{eq38} H_{y} = \frac{\dot{B}}{B} = \frac{3\ell}{2(m+1)}{T}^{-1} + \frac{1}{4}c_{4}
(T)^{-\frac{3}{n}} \;,
\end{equation}
\begin{equation}
\label{eq39} H_{y} = \frac{\dot{B}}{B} = \frac{3\ell}{2(m+1)}{T}^{-1} - \frac{1}{4}c_{4}
(T)^{-\frac{3}{n}}.
\end{equation}
The Hubble parameter, expansion scalar and shear of the model are respectively given by
\begin{equation}
\label{eq40} H = \ell(T)^{-1}  ,
\end{equation}
\begin{equation}
\label{eq41} \theta = 3\ell(T)^{-1} ,
\end{equation}
\begin{equation}
\label{eq42} \sigma^{2} = \frac{3\ell^{2}(2m-1)^{2}}{2(m+1)^{2}}{T}^{-2}
+ \frac{1}{4}c_{4}^{2}{T}^{-\frac{6}{n}}.
\end{equation}

The spatial volume ($V$) and anisotropy parameter $(\bar{A})$ are found to be
\begin{equation}
\label{eq43} V = {T}^{\frac{3}{n}},
\end{equation}
\begin{equation}
\label{eq44}  \bar{A} = \frac{1}{3}\sum_{i = 1}^{3}{\left(\frac{\triangle H_{i}}{H}\right)^{2}} =
\frac{(2m-1)^{2}}{(m+1)^{2}} + \frac{c_{4}^{2}}{6\ell^{2}} {T}^{\frac{2(n-3)}{n}},
\end{equation}
where $\triangle H_{i} = H_{i} - H (i = 1, 2, 3)$. \\\\
From the above results, it can be seen that the spatial volume is zero at $T = 0$,
and it increases with the cosmic time. The parameters $H_{i}$, H, $\theta$ and $\sigma$ diverge at the
initial singularity. The mean anisotropic parameter is an increasing function of time for $n > 3$ whereas
for $n < 3$ it decreases with time. Thus, the dynamics of the mean anisotropy parameter depends on the
value of n. Since $\frac{\sigma^{2}}{\theta^{2}} = $ constant (from early to late time), the model does
not approach isotropy through the whole evolution of the universe
\subsection{String Cosmology with Exponential-law}
Solving the equations (\ref{eq17}), (\ref{eq22}) and (\ref{eq24}), we
obtain the metric functions as
\begin{equation}
\label{eq45} A(t) = c_{2}^{\frac{3m}{m+1}}\exp{\left(\frac{3m\ell}{m+1}t\right)}\; ,
\end{equation}
\begin{equation}
\label{eq46} B(t) = \sqrt{c_{3}}c_{2}^{\frac{3}{2(m+1)}}\exp{\left[\frac{3\ell}{2(m+1)}t - \frac{c_{4}}
{6\ell c_{2}^{3}}e^{-3\ell t}\right]},
\end{equation}
\begin{equation}
\label{eq47} C(t) = \frac{c_{2}^{\frac{3}{2(m+1)}}}{\sqrt{c_{3}}}
\exp{\left[\frac{3\ell}{2(m+1)}t + \frac{c_{4}}{6\ell c_{2}^{3}}e^{-3\ell t}\right]}.
\end{equation}
\begin{figure}[ht]
\centering
\includegraphics[width=8cm,height=8cm,angle=0]{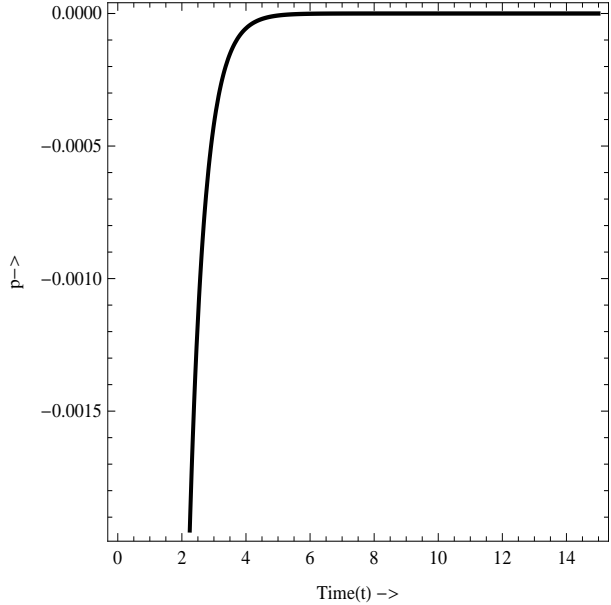} \\
\caption{The plot of pressure $p$ versus $t$ in exponential-law expansion for $q = - 1$.}
\end{figure}
\begin{figure}[ht]
\centering
\includegraphics[width=8cm,height=8cm,angle=0]{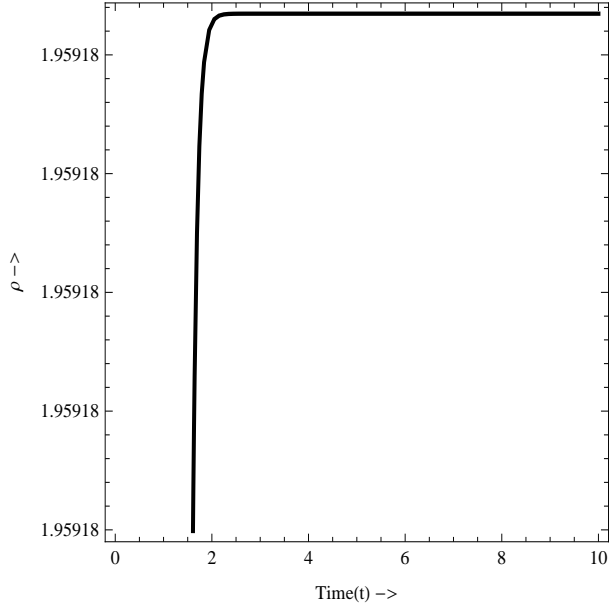} \\
\caption{The plot of energy density $\rho$ versus $t$ in exponential-law expansion for $q = -1$.}
\end{figure}
\begin{figure}[ht]
\centering
\includegraphics[width=8cm,height=8cm,angle=0]{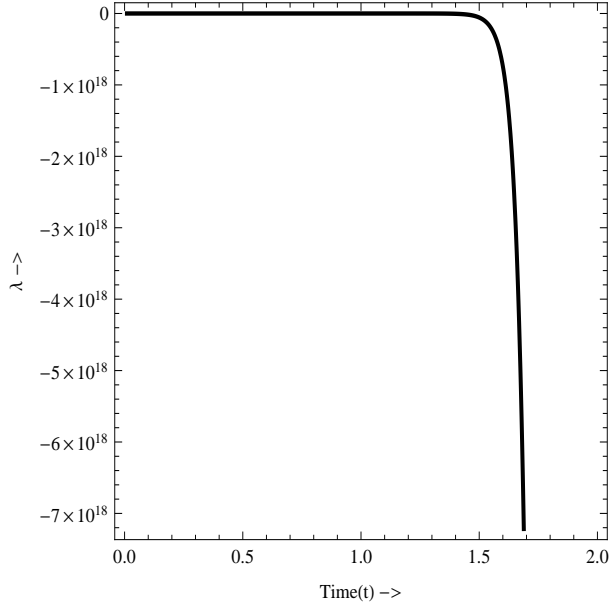} \\
\caption{The plot of string tension density $\lambda$ versus $t$ in exponential-law expansion for $q = -1$.}
\end{figure}
\begin{figure}[ht]
\centering
\includegraphics[width=8cm,height=8cm,angle=0]{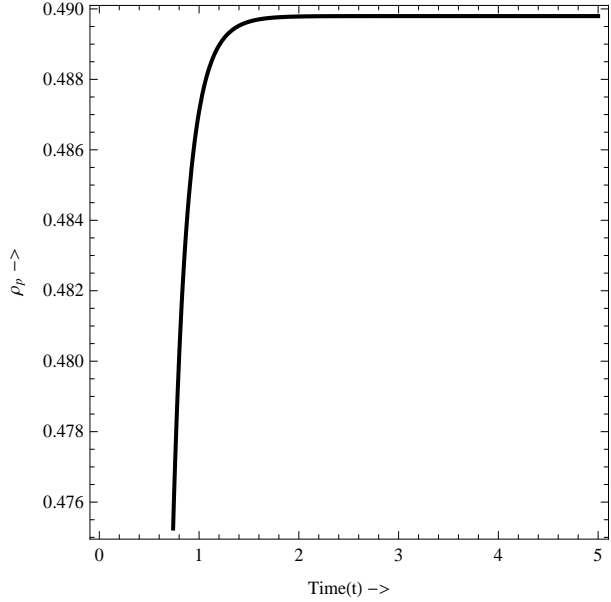} \\
\caption{The plot of particle density $\rho_{p}$ versus $t$ in exponential-law expansion for $q = - 1$.}
\end{figure}

Therefore the metric (\ref{eq1}) reduces to
\[
 ds^{2} = -dt^{2} + c_{2}^{\frac{6m}{m+1}}\exp{\left(\frac{6m\ell}{m+1}t\right)}(dx - zdy)^{2} +
\]
\[
 c_{3} c_{2}^{\frac{3}{(m+1)}}\exp{\left[2\left\{\frac{3\ell}{2(m+1)}t - \frac{c_{4}}
{6\ell c_{2}^{3}}e^{-3\ell t}\right\}\right]}dy^{2}
\]
\begin{equation}
\label{eq48}  + \frac{c_{2}^{\frac{3}{(m+1)}}}{c_{3}}
\exp{\left[2\left\{\frac{3\ell}{2(m+1)}t + \frac{c_{4}}{6\ell c_{2}^{3}}e^{-3\ell t}\right\}\right]} dz^{2}.
\end{equation}
Eq. (\ref{eq16}) gives either $\beta = 0$ or $\frac{3}{2} \dot{\beta} + \frac{3}{2} \beta \left(\frac{\dot{A}}{A} 
+ \frac{\dot{B}}{B} + \frac{\dot{C}}{C}\right) = 0$.
Therefore
\begin{equation}
\label{eq49} \frac{\dot{\beta}}{\beta}   = - \left(\frac{\dot{A}}{A} + \frac{\dot{B}}{B} + \frac{\dot{C}}{C}\right),
\end{equation}
which reduces to
\begin{equation}
\label{eq50} \frac{\dot{\beta}}{\beta} = - 3\ell.
\end{equation}
Integrating Eq. (\ref{eq50}), we obtain
\begin{equation}
\label{eq51} \beta = k_{1}e^{-3\ell t},
\end{equation}
where $k_{1}$ is an integrating constant. From Eq. (\ref{eq51}), it is observed 
that for $k_{1} > 0$ and $\ell > 0$, the displacement vector $\beta(T)$ is a decreasing function of time and it 
approaches to a small positive value at late time (i.e. present epoch), which is corroborated with Halford 
\cite{ref4} as well as with the recent observations of SNe Ia. Recent cosmological observations of SNe Ia 
(Garnavich et al. \cite{ref90,ref91}; Perlmutter et al. \cite{ref92}$-$\cite{ref94}; Riess et al.\cite{ref95,ref96}; 
Schmidt et al. \cite{ref97}). \\\\
The expressions for the isotropic pressure ($p$), the proper energy density ($\rho$), the string tension
($\lambda$) and the particle density ($\rho_{p}$) for the model (\ref{eq48}) are obtained as
\begin{equation}
\label{eq52}
p = -\frac{9\ell^{2}(4m^{2} + 2m + 1)}{4(m+1)^{2}}-\frac{c_{4}^{2}}{4c_{2}^{6}}e^{-6\ell t}
- \frac{1}{4}\exp{\left[-\frac{6\ell(1 - m)}{m + 1}t\right]} - \frac{3}{4}k^{2}_{1}e^{-6\ell t},
\end{equation}
\begin{equation}
\label{eq53}
\rho = \frac{9\ell^{2}(4m + 1)}{4(m + 1)^{2}} - \frac{c_{4}^{2}}{4c_{2}^{6}}e^{-6\ell t}
- \frac{1}{4}\exp{\left[-\frac{6\ell(1 - m)}{m + 1}t\right]} +  \frac{3}{4}k^{2}_{1}e^{-6\ell t},
\end{equation}
\begin{equation}
\label{eq54} \lambda = \frac{9\ell^{2}(1 - 2m)}{2(m + 1)} - \exp{\left[-\frac{6\ell(1 - m)}{m + 1}t\right]},
\end{equation}
\begin{equation}
\label{eq55}
\rho_{p} = \frac{9\ell^{2}(4m^{2} + 6m - 1)}{4(m + 1)^{2}} + \frac{3}{4}\exp{\left[\frac{6\ell
(m - 1)}{(m + 1)}t\right]} - \frac{c_{4}^{2}}{4c_{2}^{6}}e^{-\ell t} + \frac{3}{4}k^{2}_{1}e^{-6\ell t}.
\end{equation}
{\bf Figure 7} depicts the variation of pressure versus time for  the exponential-law evolution of the universe. We
observe that the pressure is negative and constant with the evolution of the universe. This is consistent with 
the well established fact that pressure is negative in the accelerating universe. \\\\
From Eq. (\ref{eq53}), it is observed that the rest energy density $\rho$ is a constant function of time and
$\rho > 0$ always. The rest energy density has been graphed versus time in {\bf Figure 8}. This means there is no
density evolution in this set up. A  possible reason for no evolution of density is that expansion 
of the Universe could be much rapid in which matter do not get time to re-adjust on expansion range or may be other 
unknown dominated effect which not incorporated in potential functions of this space-time for exponential-law.\\\\
From Eq. (\ref{eq54}), it is observed that the tension density $\lambda$ is a  decreasing function of time and
$\lambda < 0$ always. {\bf Figure 9} shows the plots of string tension density verses time However, initially it 
remains constant and then decreases more sharply with the cosmic time. \\\\
{\bf Figure 10} shows the plots of particle density versus time. In early stage of evolution, it increases with 
time and later stage it remains constant. 
From Eq. (\ref{eq55}), it is evident that the particle density $\rho_{p}$ is an increasing function of time and 
$\rho_{p} > 0$ for all time. \\\\
The expressions for kinematical parameters i.e. the scalar of expansion ($\theta$), shear scalar ($\sigma$),
the spatial volume ($V$), average anisotropy parameter ($\bar{A}$) and deceleration parameter ($\rm q$) for
the model (\ref{eq48}) are given by
\begin{equation}
\label{eq56} \theta = 3 \ell,
\end{equation}
\begin{equation}
\label{eq57} \sigma^{2} = \frac{3\ell^{2}(2m-1)^{2}}{4(m+1)^{2}}-\frac{c_{4}^{2}}{4c_{2}^{6}}e^{-6\ell t},
\end{equation}
\begin{equation}
\label{eq58} V = c_{2}^{3}e^{3\ell t},
\end{equation}
\begin{equation}
\label{eq59} \bar{A} = \frac{(2m-1)^{2}}{2(m+1)^{2}} - \frac{c_{4}^{2}}{6\ell^{2}c_{2}^{6}}e^{-6\ell t},
\end{equation}
\begin{equation}
\label{eq60} q = - 1.
\end{equation}
The rotation $\omega$ is identically zero. The negative value of $q$ indicates inflation. The evolution of
the universe in such a scenario may not be consistent with the present day observations predicting
an accelerated expansion.   \\\\
The rate of expansion $H_{i}$ in the direction of x, y and z are given by
\begin{equation}
\label{eq61} H_{x} =  \frac{3m\ell}{m + 1},
\end{equation}
\begin{equation}
\label{eq62} H_{y} =  \frac{3\ell}{2(m + 1)} + \frac{c_{4}}{2c_{2}^{3}}e^{-3\ell t},
\end{equation}
\begin{equation}
\label{eq63} H_{z} =  \frac{3\ell}{2(m + 1)} -\frac{c_{4}}{2c_{2}^{3}}e^{-3\ell t}.
\end{equation}
Hence the average generalized Hubble's parameter is given by
\begin{equation}
\label{eq70} H = \ell.
\end{equation}
From equations (\ref{eq61})-(\ref{eq63}), we observe that the directional Hubble parameters are time
dependent while the average Hubble parameter is constant. \\\\
It is observed that the physical and kinematic quantities are all constant at $t = 0$ as well as $t \to
\infty$. The expansion in the model is uniform throughout the time of evolution. We find that
$\lim_{t \rightarrow 0}\frac{\sigma^{2}}{\theta^{2}} \ne  0$ and $\lim_{t \rightarrow \infty}
\frac{\sigma^{2}}{\theta^{2}} \ne  0$, which show that this inflationary universe eventually does not
approach isotropy for early as well as large values of $t$. The derived model is non-singular.
\section {Conclusions}
In this paper we have obtained a new class of spatially homogeneous and anisotropic Bianchi-II cosmological 
models in presence of perfect fluid distribution of matter representing massive strings within the framework of normal 
gauge in Lyra's manifold by applying the variation law for generalized Hubble's parameter that yields a constant value 
of deceleration parameter in general relativity. Secondly we have also assumed that the expansion in the model is 
proportional to the shear as Collins et al. \cite{ref78} have showed that the normal congruence to the homogeneous 
expansion satisfies that the condition $\frac{\sigma}{\theta} = $ constant. The law of variation for Hubble's 
parameter defined in (\ref{eq19}) for B-II space-time models give two types of cosmologies, (i) first form 
(for $n \ne 0$) shows the solution for positive value of deceleration parameter for ($n > 1$) and negative value 
of deceleration parameter for ($n < 1$) indicating the power law expansion of the universe whereas (ii) second one 
(for $ n = 0$ shows the solution for negative value of deceleration parameter, which shows the exponential expansion 
of the universe. Exact solutions of Einstein's field equations for these models of the universe have been obtained
by using the two forms of average scale factor. The power law solutions represent the singular model where
the spatial scale factors and volume scalar vanish at $T = 0$. The energy density and pressure are infinite
at this initial epoch. As $T \to \infty$, the scale factors diverge and $p$, $\rho$, $\lambda$ and $\rho_{p}$ tend 
to zero. $A_{m}$ and $\sigma^{2}$ are very large at initial time but decrease with cosmic time and vanish as 
$T \to \infty$. The exponential solutions represent singularity free model of the universe. In this case as 
$T \to - \infty$, the scale factors tend to zero which indicates that the universe is infinitely old and has 
exponential inflationary phase. All the parameters such as scale factor, $p$, $\rho$, $\lambda$, $\rho_{p}$, $\theta$, 
$\sigma^{2}$ and $A_{m}$ are constant at $t = 0$. \\\\
Under the law of variation for Hubble's parameter defined in (\ref{eq19}), it has been shown that the two
classes of solutions lead to the conclusion that, if $q > 0$ the model expands but always decelerate whereas
$q < 0$ provides the exponential expansion and later accelerates the universe. The evolution of the universe in
such a scenario is shown to be consistent with the present observations predicting an accelerated expansion. \\\\
The models presents the dynamics of strings in the accelerating and decelerating modes of evolution of the universe. 
It has been found that massive strings dominate in the decelerating universe whereas strings dominate in the 
accelerating universe. The strings dominate in the early universe and eventually disappear from the universe 
for sufficiently large times. This is in consistent with the current observations. \\\\
It is observed that the displacement vector $\beta(t)$ matches with the nature of the cosmological constant 
$\Lambda$  which has been supported by the work of several authors as discussed in the physical behaviour of 
the models in Subsections $3.1$ and $3.2$. In both cases (3.1) and (3.2) the the displacement vector $\beta$ 
is a decreasing function of time and it approaches to a small positive value at late time (i.e. present epoch), 
which is corroborated with Halford \cite{ref4} as well as with the recent observations of SNe Ia (Garnavich et al. 
\cite{ref90,ref91}; Perlmutter et al. \cite{ref92}$-$\cite{ref94}; Riess et al. \cite{ref95,ref96}; Schmidt et al. 
\cite{ref97}).  In recent past there is an upsurge of interest in scalar fields in general relativity and 
alternative theories of gravitation in the context of inflationary cosmology \cite{ref100,ref101,ref102}. Therefore 
the study of cosmological models in Lyra's manifold may be relevant for inflationary models. There seems a good 
possibility of Lyra's manifold to provide a theoretical foundation for relativistic gravitation, astrophysics and 
cosmology. However, the importance of Lyra's manifold for astrophysical bodies is still an open question. In fact, 
it needs a fair trial for experiment. \\\\
It is remarkable to mention here that our work generalizes the results recently obtained by Kumar \cite{ref103} in 
2010. In absence of displacement field $\phi_{i}$ (i.e. by putting $k = 0$ in case (3.1)), we obtain the results 
obtained by Kumar \cite{ref103}.
\section*{Acknowledgements}
\noindent
One of the authors (A. Pradhan) would like to thank the Laboratory of Information Technologies, Joint Institute 
for Nuclear Research, Dubna, Russia for providing facility and hospitality where part of this work was done.
\newline

\end{document}